\documentclass [submission,copyright,creativecommons]{eptcs}
\providecommand{\event}{GraMSec 2014} 
\usepackage{graphicx}

\title{Towards Automating the Construction \& Maintenance of Attack Trees: a Feasibility Study}
\author{St\'ephane Paul
\institute{Thales Research \& Technology\\ Palaiseau, France}
\email{stephane.paul@thalesgroup.com}
}
\def\titlerunning{Technique for Automating the Construction \& Maintenance of Attack Trees}
\def\authorrunning{St\'ephane Paul}
\begin{document}
\maketitle

\begin{abstract}
Security risk management can be applied on well-defined or existing systems; in this case, the objective is to identify existing vulnerabilities, assess the risks and provide for the adequate countermeasures. Security risk management can also be applied very early in the system's development life-cycle, when its architecture is still poorly defined; in this case, the objective is to positively influence the design work so as to produce a secure architecture from the start. The latter work is made difficult by the uncertainties on the architecture and the multiple round-trips required to keep the risk assessment study and the system architecture aligned. This is particularly true for very large projects running over many years. This paper addresses the issues raised by those risk assessment studies performed early in the system's development life-cycle. Based on industrial experience, it asserts that attack trees can help solve the human cognitive scalability issue related to securing those large, continuously-changing system-designs. However, big attack trees are difficult to build, and even more difficult to maintain. This paper therefore proposes a systematic approach to automate the construction and maintenance of such big attack trees, based on the system's operational and logical architectures, the system's traditional risk assessment study and a security knowledge database.
\end{abstract}

\section{Introduction}

Large industries perform security risk assessments for large new systems whose development life-cycles span over many years. The security risk assessment work is to begin very early in the development life-cycle of the systems so as to positively influence the design towards secure architectures; the work is difficult because it has to be run on a yet poorly defined system architecture and because the risk assessment has to consider the complete system, i.e. an aggregation of premises, equipment, people and procedures. The amount of work justifies that multiple security experts are involved on the project, possibly from multiple sites and background. In this context, traditional risk assessment approaches, usually based on ISO-31000  \cite{iso:31000} and ISO-27001 \cite{iso-iec:27001}, reach their limit, in particular with respect to human cognitive scalability.
On the European Galileo programme, the use of graphical attack trees has been shown to provide a positive contribution to traditional risk assessment approaches. However, the construction and maintenance of attack trees are tedious and error-prone tasks. Moreover, in a multi-user context, construction approaches differ from one security expert to another, which renders the reading and maintenance of attack trees even more complex for third-parties.

This paper reports on a feasibility study that was run to assess if it is possible to automate the construction \& maintenance of attack trees using yet poorly defined architectural artefacts. This paper proposes a preliminary layer-per-layer systematic approach to generate skeletons of attack trees based on information coming from the system's architecture, as classically established using an architecture framework (AF), and information coming from the system's security risk assessment study, as traditionally run using a given risk assessment method and its related security knowledge base, e.g. EBIOS \cite{ebios}.  
The support of an industrially used AF was a very important constraint imposed on this feasibility study; indeed, we wanted to assess how much of an attack tree could be generated without additional work from system architects and/or designers. This feasibility study was run using empirical data from the Galileo risk assessment programme (cf. chapter 2 below), but obvious confidentiality reasons do not permit us to publish such data. Thus, this paper details the steps of the approach, and illustrates it on a running example from the automotive domain, which is public and easily understandable by all.

\section{Scientific and Empirical Baseline to Defining the Approach}

Galileo is Europe's own global navigation satellite system, providing a highly accurate, guaranteed global positioning service under civilian control. It is inter-operable with GPS and GLONASS, i.e. the US and Russian global satellite navigation systems. It corresponds to a system based on 30 satellites, 2 main control centres and 25 unmanned world-wide sites. The system is developed by the European Space Agency (ESA) for the European Commission (EC).
To satisfy the security objectives identified in the frame of the Galileo programme, it was requested to define a risk management process. This needed to be an evolving process, taking into account the global life cycle of the system, from its specification to its operation. The main objective of the process to be defined was to provide the security accreditation authority assurance that the Galileo system is secure enough for authorising: (i) the launches of the satellites; (ii) the deployment of the ground infrastructure; (iii) the operation of the Galileo services.

Considering the above, the risk management process was defined through a series of brainstorming sessions, involving 10 experts from EC, ESA and industry, over a period of approximately 6 months. Techniques used in cyber-security and safety were analysed with respect to their applicability at system-level, whereby the system comprises equipment, people and procedures. Initial risk management process proposals were consolidated by assessing their direct application on the Galileo programme itself. The process was finally approved by the 27 Member States, in Sept. 2011, just before the first satellite launch. It is now proposed as the reference risk management process for other major EC programmes, e.g. EGNOS.

The graphical representation of risks on attack trees permitted the security accreditation authority to: (i) gain an intuitive approach of the risk, (ii) associate each risk to the Galileo System Design; and (iii) better perceive the impact of risk treatment on the system architecture.
The success of the process defined for the Galileo programme has conducted Thales to generalise the risk management process for all IT systems under its responsibility.
The approach proposed in this document is in a large part based on the empirical background gained through the Galileo programme. It also rests upon an important state of the art analysis \cite{DBLP:journals/corr/abs-1303-7397} \cite{62441619-179/7}. It will be further consolidated on other commercial risk assessment studies.

\section{The Running Example}

The running example is based on the \emph{Loss of Integrity the Manual Breaking capability in a standard modern Car used as Taxi}.  The running example has been modelled using the Thales Melody AF. It is a simplistic model; the objective here is not to build a realistic car model, but to include the key modelling artefacts that can be used to generate an attack tree.

\begin{figure}[htbp]
\begin{center}
\includegraphics[width=10cm]{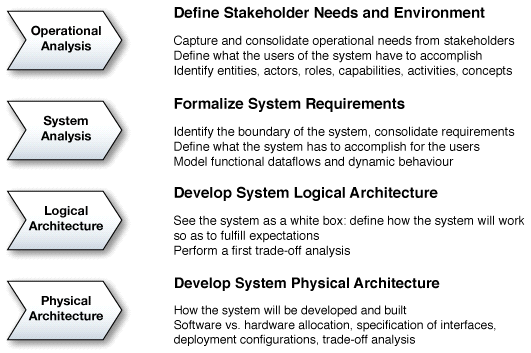}
\caption{The four abstraction levels of the Thales Melody framework}
\label{melody}
\end{center}
\end{figure}

The Thales Melody AF supports architecting using four abstraction levels, as illustrated in Figure \ref{melody}. Under our assumption that the security risk assessment is being performed early in the system's development life-cycle, the  {\tt Physical} abstraction level of Melody has not been used in our approach.

The Thales Melody Architecture Frameworks is UMLish. It is our believe that most state-of-the-art and commercially available AFs would provide similar capabilities to support the proposed approach, so alternative AFs could have been used, but a detailed assessment has not (yet) been performed.

\section{Attack Tree Construction Overview}

Our approach uses the dominant type of attack tree model, i.e. the Boolean-logical tree based approach in which the top or root node of the tree represents a defender's feared event.
Given a feared event, our proposal is to systematically structure the attack tree in layers according to the following elements:
\begin{itemize}
\item system states and modes, in which the occurrence of the feared event makes sense;
\item supporting asset types (e.g. hardware, software) that could potentially be attacked;
\item attack entry points (i.e. supporting asset interfaces) for each supporting asset type;
\item threats that can be exercised on the attack entry points;
\item threat sources that can exercise the threats.
\end{itemize}
The two high-level principles governing the tree construction are the following:
\begin{itemize}
\item the feared events are defined at strategic level: they are driven by operational considerations,
\item logical {\tt AND} gate decompositions should be located as low as possible in the tree.
\end{itemize}

The first principle originates from the customer and enforces a true risk-based approach (by opposition to technical security). Moreover, when the feared event at the root of an attack tree is driven by operational considerations, it is simple to assign an individual stakeholder to the attack tree, with which the security expert will be able to discuss the relevance and completeness of the attack tree decomposition.

The second principle is driven by attack tree exploitation and ergonomic considerations. Conjunctive Boolean gates create dependencies between tree branches. Locating {\tt AND} gates as low as possible in the tree limits the span of dependent branches and increases readability of the attack trees.
In this set-up, {\tt AND} gates are typically used to capture:
\begin{itemize}
\item pre-conditions to attack enacting, for example, knowledge about the supporting asset, or, more complex, a change in state and mode;
\item conditions to make succeed the attack, in particular with respect to system redundancy;
\item post-conditions, typically to allow for the repudiation of the attack. 
\end{itemize}

\section{Step-by-Step Description of the Approach}

\subsection{Step 1: Creation of the Attack Tree Root}

Feared events are classically captured textually in the security risk assessment study. For our running example, let us suppose that the feared event is defined as the \emph{Loss of Integrity of the Manual Braking operational process on the Car}. Thus, it is easy to extract the feared event from the security risk assessment study to create the attack tree root. The technical challenge here is to map this informal statement with artefacts of the system's architecture and security knowledge base.

\begin{figure}[htbp]
\begin{center}
\includegraphics[width=15cm]{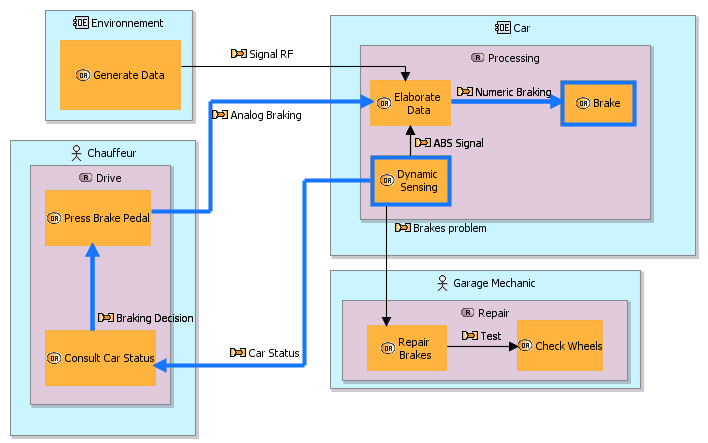}
\caption{Running example - the manual braking operational process}
\label{opsProcess}
\end{center}
\end{figure}

From the system's operational architecture (cf. Figure \ref{opsProcess}) it is possible to recognise the {\tt Car} as being an Operational Entity (OE). The {\tt Manual Braking} operational process is pictured as a sequence of operational activities (OAs); the first activity (i.e. {\tt Dynamic Sensing}) and the last activity (i.e. {\tt Brake}) of the operational process are surrounded by thick borders; the interactions between the operational activities are pictured by thick arrows; other (thin) arrows represent interactions that are not part of the {\tt Manual Braking} operational process.

From our security knowledge base, it is possible to recognise the term {\tt Integrity} as being a standard security criterion; {\tt Confidentiality} and {\tt Availability} would likewise be eligible.

Our proposal is not to perform automatic recognition on informally stated feared events, but rather to impose a strict grammar to the description of feared events in the security risk assessment study, typically:

\begin{center}
Loss of [\emph{security criterion}] of the [\emph{primary asset}] on the [\emph{operational entity}].
\end{center}

To capture the security criterion, the security expert would select an option from a drop down menu, or similar mechanism, in line with the security knowledge base selected for the study.
To capture the operational entity (i.e. the {\tt Car}), the security expert would select the artefact directly from the system's operational architecture by browsing through the list of defined operational entities. Such an approach has already been successfully experimented in a security risk assessment context \cite{SCD4-4b}.
In the above construct, the {\tt Manual Braking} operational process of the AF is not referenced directly. Instead, we have used the concept of \emph{primary asset}, as defined in the French EBIOS risk assessment method, i.e. something that is of value \cite{ebios}. It is assumed that the primary asset has been previously mapped to the operational process artefact in the system's operational architecture (i.e., the {\tt Manual Braking} operational process, in a similar manner as for the operational entity above \cite{SCD4-4b}.

The severity of the feared event originates from the risk assessment study. In our running example, it is assumed to be {\tt Critical}.

\subsection{Step 2: Structuring the Tree According to States and Modes}

The second step in constructing the attack tree is to take into consideration the system's states and modes in which the realisation of the considered feared event makes sense. The Melody AF allows for a \emph{state machine X operational activities} matrix, which defines in which states and modes the operational processes can be run. From this matrix, the attack tree can be generated as follows:
\begin{itemize}
\item a first decomposition of the feared event is made with all the states in which the operational process can run, connected with an {\tt OR} gate; 
\item a decomposition of the state nodes can then be made with all the modes in which the operational process can run, connected with an {\tt OR} gate to the corresponding state node in the tree.
\end{itemize}

\begin{figure}[htbp]
\begin{center}
\includegraphics[width=12cm]{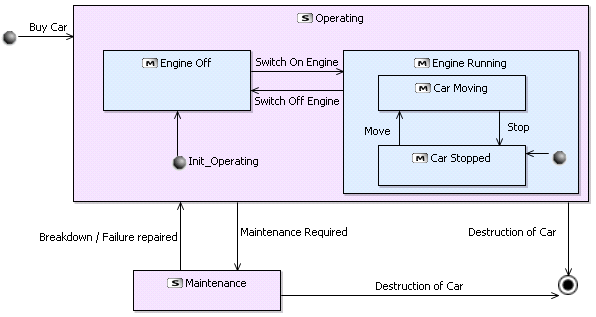}
\caption{Running example - the manual braking operational process}
\label{SandM}
\end{center}
\end{figure}

\begin{figure}[htbp]
\begin{center}
\includegraphics[width=8cm]{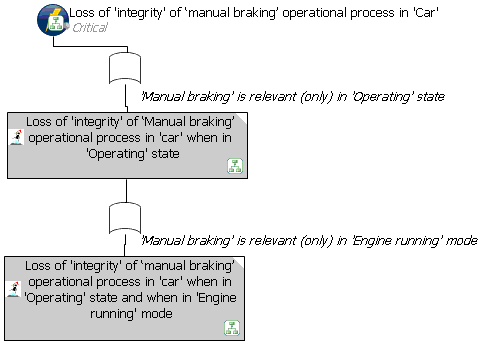}
\caption{Automatically constructed attack tree based on states and modes}
\label{SandMtree}
\end{center}
\end{figure}

For our running example, let us consider a simple state machine, as illustrated in Figure \ref{SandM}. Let us also make the reasonable assumption that the {\tt Manual Braking} operational process can be run only in the {\tt Operating} state and in the {\tt Engine Running} mode of the {\tt Car}. Thus, in the attack tree, there is need for only one branch corresponding to the {\tt Operating} state, and there is need for only one branch corresponding to the {\tt Engine Running} mode. The resulting tree is shown in Figure \ref{SandMtree}.

It is noteworthy that the labelling of the attack tree nodes can be automatically and dynamically generated based on the system's operational architecture. Dynamicity stems from the traceability links that are established between the tree nodes and the states and modes in the operational architecture. Thus, if the name of a state or mode changes, the label of the corresponding node in the attack tree can be automatically and immediately updated. Likewise, if a state or mode is deleted, a warning can be attached to the corresponding subtree, calling for specific attention by the security expert. 
The running example illustrated here is obviously simplistic, but it is sufficient to capture the general idea that an attack depends on the states and modes of the targeted system.

It is worthwhile recalling here that elaborating the attack tree with states and modes is not just a tree structuring feature. States and modes carry semantics about system behaviour, and therefore also about what attacks are possible at a given time. In step 6 (see below), we introduce attack preconditions; some of these conditions can be a change of system state or mode, which inherently transforms the attack tree into a Directed Acyclic Graph (DAG) or creates duplicate sub-trees.

\subsection{Step 3: Structuring the Tree According to Supporting Asset Types}

Security knowledge bases often organise supporting assets in types in order to define categories of vulnerabilities with their corresponding threats, e.g. paper documents can burn, hardware can be stolen, networks can be saturated, people can be influenced. In this 3rd step of the attack tree construction, we make use of this supporting asset typology in order to map the attack tree structure with the knowledge base. The goal here is to increase the readability of the attack tree and assure its completeness with respect to best practices as captured in the security knowledge bases. However, this structuring is optional and can be skipped if felt as unnecessary.

For our running example, let us suppose the existence of four supporting asset types: {\tt Hardware}, {\tt Software}, {\tt Networks} and {\tt Organisations}. At this very early stage in the architectural design, {\tt Network} must not be understood as routers, servers, switches, and the like, but rather as functional and / or component exchanges, including social networks. Focus is more on the data that is exchanged in support of the operational processes, than on the equipment that supports the data exchanges.

\begin{figure}[htbp]
\begin{center}
\includegraphics[width=16cm]{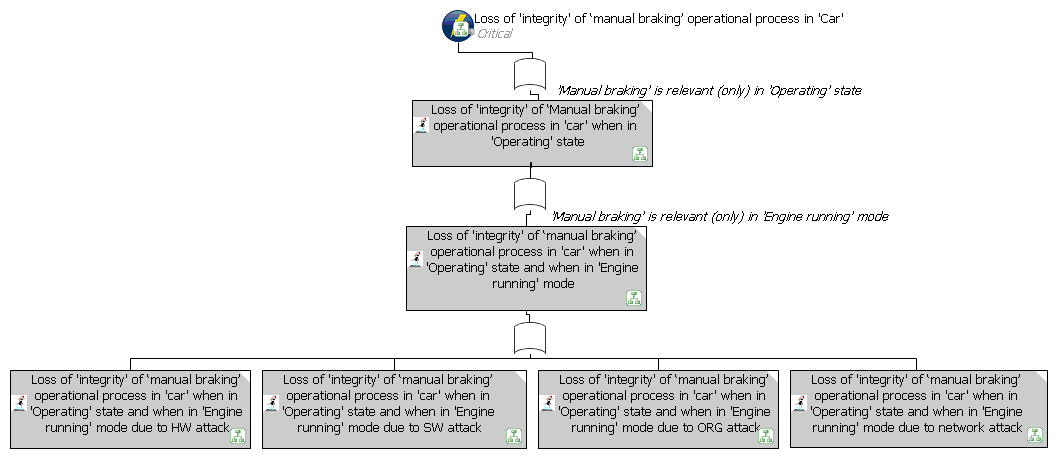}
\caption{Automatically constructed attack tree based on supporting asset types}
\label{SATtree}
\end{center}
\end{figure}

The resulting tree is shown in Figure \ref{SATtree}.
As in the previous step, the labelling of the attack tree nodes can be automatically generated based on the content of the security knowledge base.

\subsection{Step 4: Structuring the Tree According to Attack Entry Points}

During the fourth step, the objective is to capture the supporting assets that can be targeted as attack entry points by the attacker. This exercise requires identifying all the possible attack entry points defined in the logical architecture, supported by the traceability links between the artefacts of the operational architecture and their corresponding elements in the logical architecture (cf. Figure \ref{melody}).

\begin{figure}[htbp]
\begin{center}
\includegraphics[width=16cm]{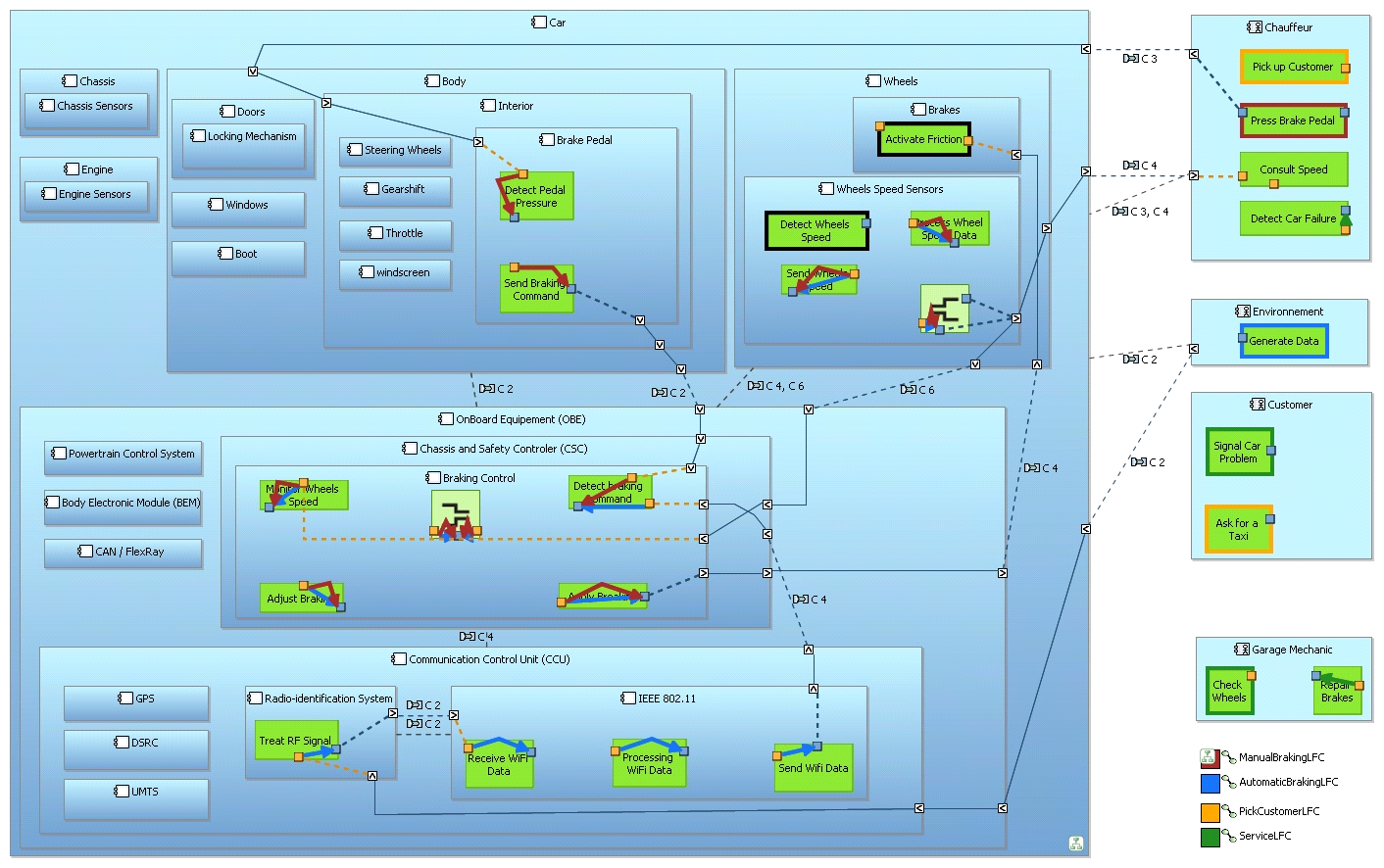}
\caption{Running example - simplified logical architecture model}
\label{LogArchi}
\end{center}
\end{figure}

\begin{figure}[htbp]
\begin{center}
\includegraphics[width=16cm]{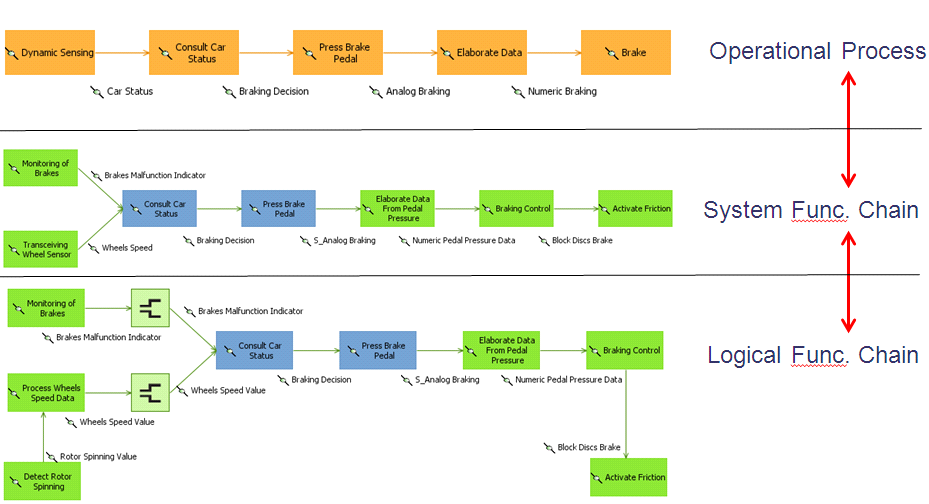}
\caption{Running example - traceability between the operational process and the logical functional chain}
\label{Traceability}
\end{center}
\end{figure}

Figure \ref{LogArchi} provides a simplified logical architecture model for our running example. In this model, a {\tt Manual Braking} functional chain has been defined that corresponds to the {\tt Manual Braking} operational process defined in the operational architecture.
The Thales Melody AF allows one to trace artefacts between abstraction layers, as illustrated for our running example in Figure  \ref{Traceability}.

\begin{figure}[htbp]
\begin{center}
\includegraphics[width=9cm]{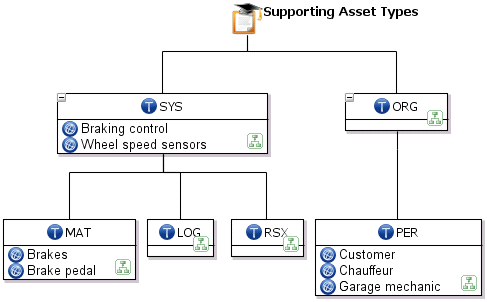}
\caption{Running example - tagging of supporting assets in the security risk assessment study}
\label{Typology}
\end{center}
\end{figure}

In the Melody AF, and presumably in most logical architecture frameworks, the logical components are not assigned tags specifying their type. However, this work is traditionally done in security risk assessment studies, as illustrated for our running example in Figure \ref{Typology}. In this figure it can be seen that:
\begin{itemize}
\item the {\tt Brake Pedal} and {\tt Brakes} (meaning here {\tt Break Pads}) are hardware components,
\item the {\tt Customer}, {\tt Chauffeur} and {\tt Garage Mechanic} are people, and
\item the {\tt Braking Control} and {\tt Wheel Speed Sensors} are complex systems, including hardware, software and / or network sub-components.
\end{itemize}

Based on all the aforementioned information, it is now possible to continue the decomposition of the attack tree: for all leaf nodes of the attack tree corresponding to {\tt Hardware}, {\tt Software} and {\tt Organisations} add the corresponding identified attack entry points, connected with {\tt OR} gates, if they are involved in the logical functional chain corresponding to the operational process referenced in the feared event at the tree root.

For {\tt Network}-related nodes, the construct is a bit more complex. The objective here is to identify all the (external) logical component interfaces that could be used as attack entry points in threatening the functional chain. The search of these relevant logical component interfaces is highly dependant on the selected Architecture Framework, and will not be detailed here. Assuming they are correctly identified, it is possible to complete the attack tree decomposition: add all the identified attack entry points related to interfaces under the {\tt Network} leaf node of the tree, connected with an {\tt OR} gate.

\begin{figure}[htbp]
\begin{center}
\includegraphics[width=15.5cm]{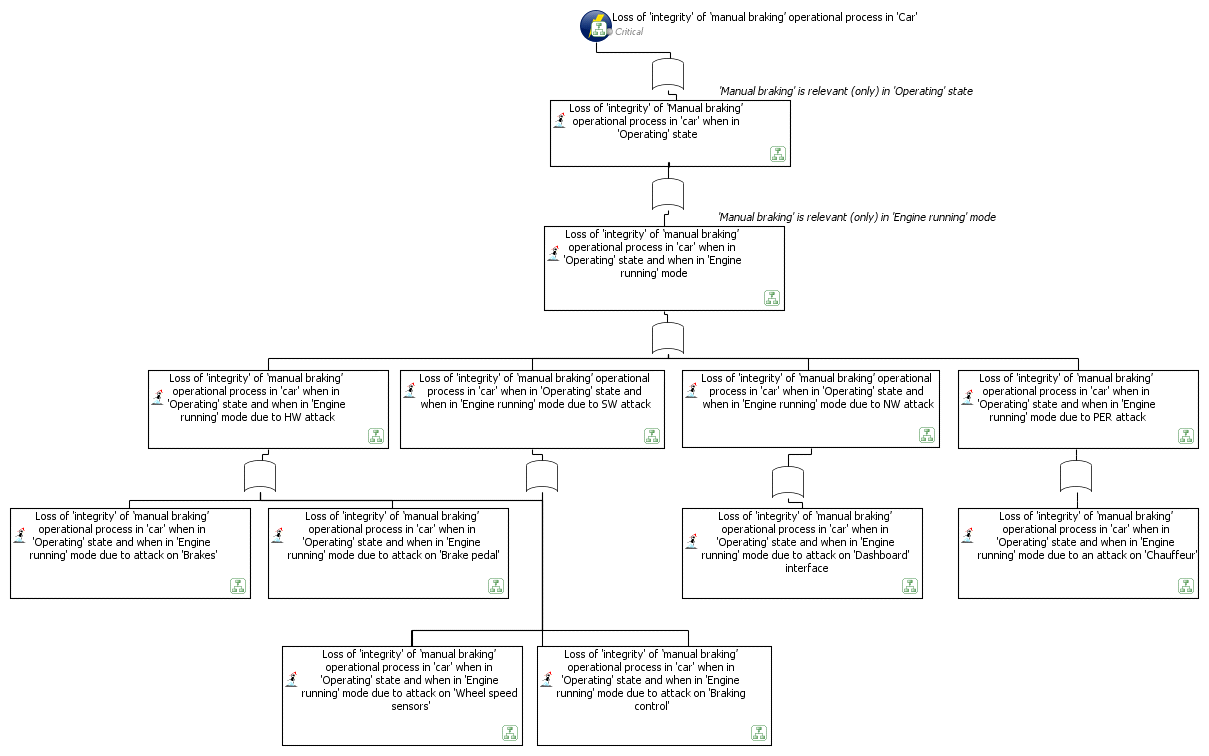}
\caption{Automatically constructed attack tree based on attack entry points}
\label{AEPtree}
\end{center}
\end{figure}

In our running example, the two hardware supporting assets can be added below the hardware attack node, connected with an {\tt OR} gate. For people, only the {\tt Chauffeur} needs to be added, because the other two persons are not involved in the logical functional chain.
For elements typed as SYS (cf. Figure \ref{Typology}), new nodes need to be created both under the {\tt Hardware} and {\tt Software} attack nodes; the construction process is similar to the one described above, but it should be noted that SYS elements lead either to the duplication of nodes in the tree, or to the creation of a directed acyclic graph (DAG) if the tooling supports such a construct. The management of the {\tt Network} components of SYS elements is explained below.
In our simple running example, we have defined two external interfaces:
\begin{itemize}
\item the {\tt Dashboard}, which allows for the display of the car's speed and also provides a malfunction indicator;
\item the {\tt Brake Pedal}, which allows the driver to apply an analogic braking force that somehow translates to a braking force on the brake pads.
\end{itemize}
Only the {\tt Dashboard} interface is concerned by the two logical components tagged as {\tt SYS}. Thus only the {\tt Dashboard} interface should be added in the attack tree under the {\tt Network} attack node. The complete resulting attack tree (in fact a DAG), at the end of this step, is illustrated in Figure \ref{AEPtree}.

\subsection{Step 5: Structuring the Tree According to Applicable Threats}

In this fifth step, we again use the security knowledge base to decompose each supporting asset leaf node of the attack tree with all realistic threats that can apply, connected with an {\tt OR} gate. We assume that the security knowledge base provides a list of threats characterised by at least the targeted supporting asset type and the concerned security criteria, as illustrated in Table \ref{threatExamples}.
The proposed algorithm systematically scans the security knowledge base selected for the study; it therefore ensures the completeness of the threat study with respect to that database.
As in the previous steps, the labelling of the attack tree nodes can be automatically generated based on the content of the security knowledge base.

\begin{table}[htdp]
\caption{Examples of threat descriptions in the EBIOS knowledge base \cite{ebios}}
\begin{center}
\begin{tabular}{|p{1cm}|p{1.8cm}|p{2cm}|p{2cm}|p{3.5cm}|p{3.5cm}|}
\hline
\textbf{Threat} & \textbf{Targeted supporting asset} & \textbf{Description} & \textbf{Concerned security criteria} & \textbf{Exploited vulnerabilities} & \textbf{Pre-requisites} \\
\hline
\textbf{MAT-MOD} & Hardware & Hardware modification & Availability, Integrity, Confidentiality & Elements can be added, retrieved or substituted; Elements can be deactivated & Knowledge of the existence and location of the hardware; Physical access to the hardware \\ \hline
\textbf{RSX-USG} & Network & Man-in-the-middle attack & Availability, Integrity & Flow content can be altered; Flow rules can be altered; Is the unique transmission resource & Knowledge of the existence and location of the canal; Physical or logical access to the canal \\ \hline
\end{tabular}
\end{center}
\label{threatExamples}
\end{table}%

According to the EBIOS knowledge base \cite{ebios}, amongst the six threats targeting people, only the {\tt Influence over a person} and {\tt Overloading of the capacity of a person} threats are related to the {\tt Integrity} security criterion. Thus, in our running example, the attack node on the {\tt Chauffeur} can be decomposed with two attack sub-nodes. All the other attack entry points are treated similarly. Due to space limitations in this paper, the resulting tree is not shown.

\subsection{Step 6: Structuring the Tree According to Threat Sources}

The sixth and last step of our algorithm is the most complex. For this step, we make the reasonable assumption  that the threat sources are explicitly defined in the risk assessment study. The complexity of this algorithm step stems from the necessity of adequately selecting the threat sources to be added in the attack tree, based on the fact that they can realistically enact the threats. Selecting unrealistic threat sources may lead the attack tree end-users to reject the attack tree because uselessly oversized; on the contrary, retaining too few threat sources may compromise the completeness of the study.

Each threat has prerequisites, in particular attack entry point access prerequisites that must be satisfied by a threat source for that threat source to be retained in the attack tree. For example, to be able to modify some hardware equipment, a threat source requires knowledge about the existence \& location of the hardware, and needs physical access to that hardware equipment (cf. Table \ref{threatExamples}).

The system's logical architecture provides information on the existence of access possibilities, typically based on the existence of logical component ports. In our running example, access to the {\tt Brake Pedal} is possible only through the {\tt Body} and {\tt Interior} of the {\tt Car}.

The system's logical architecture also provides information on the expected uses of these access possibilities, typically based on functional chains. In our running example, Figure \ref{LogArchi} shows that access to the {\tt Brake Pedal} is performed as part of the {\tt Manual Braking} functional chain.

Finally, the system's logical architecture provides information on the expected users of these access possibilities, typically based on operational actors Ð some of which may be considered as threat sources. In our running example, Figure  \ref{LogArchi} shows that the {\tt Chauffeur} is an operational entity that is expected to access the {\tt Brake Pedal} as part of the {\tt Manual Braking} functional chain. If the {\tt Chauffeur} is expected to use the {\tt Brake Pedal}, then it can be assumed that he has knowledge about the existence and location of the {\tt Brake Pedal}; he may even have received training about it.
Thus, if a threat source is part of the system's logical architecture (i.e. insider attacker), some attack entry point access preconditions can be checked.

However, all preconditions cannot be checked, even when the threat source is part of the system's logical architecture. For example, the {\tt Taxi Customer} has access to the {\tt Car Interior} but he is not expected to use the {\tt Brake Pedal}, so, based on the logical architecture, no assumption can be made on his knowledge of the existence and localisation of the {\tt Brake Pedal}.
Indeed, a system design expresses what a system is or can do, not what it is not or cannot do. This is a major limitation in using architectural artefacts for building attack trees. This limitation can be overcome by requesting additional expert input, but such additional inputs must be kept as low as possible to preserve the usability of the approach.

Due to paper length limitations, the conditions determining if a threat source should or should not be retained for insertion in the attack tree cannot be discussed herein. This paper however exposes the two main issues that render quite complex this step of the automation approach. It also provides the high-level algorithm and illustrates the resulting tree with the running example.

There are two main issues in selecting threat sources for insertion in the attack tree: (1) it is required to distinguish between physical and logical access pre-conditions; (2) access prerequisite satisfaction, for a known threat source, depends on the system's state and mode.

The proposed high-level algorithm to decide whether to retain or reject a threat source is the following:
\begin{itemize}
\item if the threat source is not represented as an actor in the architecture, then the threat source is retained because we lack knowledge about it: the security expert is requested to further manually develop or close this branch of the attack tree;
\item else if the threat source has the required access (physical and/or logical, as required for the attack) to the attack entry point in the considered system's state \& mode then the threat source is retained, for an intentional or an accidental attack; the security expert is requested to further manually develop or close this branch of the attack tree;
\item else if the threat source is malevolent, the threat source is retained, even though there is a doubt about its access capabilities; the security expert is requested to further manually develop or close this branch of the attack tree;
\item else (i.e. known non-malevolent threat source with no known access to the supporting asset), the threat source is rejected.
\end{itemize}
If a threat source is retained to be added in the attack tree for a given attack, then an {\tt AND} gate is added with:
\begin{itemize}
\item one or more leaf nodes representing the attack preconditions, as provided in the security knowledge base (cf. Table \ref{threatExamples});
\item the attack itself; the security expert is requested to further manually develop this attack node;
\item optionally, attack post-conditions, typically to ensure attack repudiation.
\end{itemize}
Amongst the most interesting preconditions are the ones involving a change of state and mode, which can be far in the past, or immediately before the attack. In our running example, activating malicious code in the {\tt Braking Control} system when the {\tt Car} is moving requires having injected that malicious code when the {\tt Car} was in the {\tt Maintenance} state. Here, the precondition is represented by a very complex subtree, but an online change of state and mode can be even more complex. Indeed, the attack is then enacted in the new state and mode, possibly with an impact on the severity, or even the relevance, of the feared event.

This algorithm step is the first in which we use the {\tt AND} logical gate, in accordance with our aim to introduce conjunctions as low as possible in the attack tree.

\begin{figure}[htbp]
\begin{center}
\includegraphics[width=16cm]{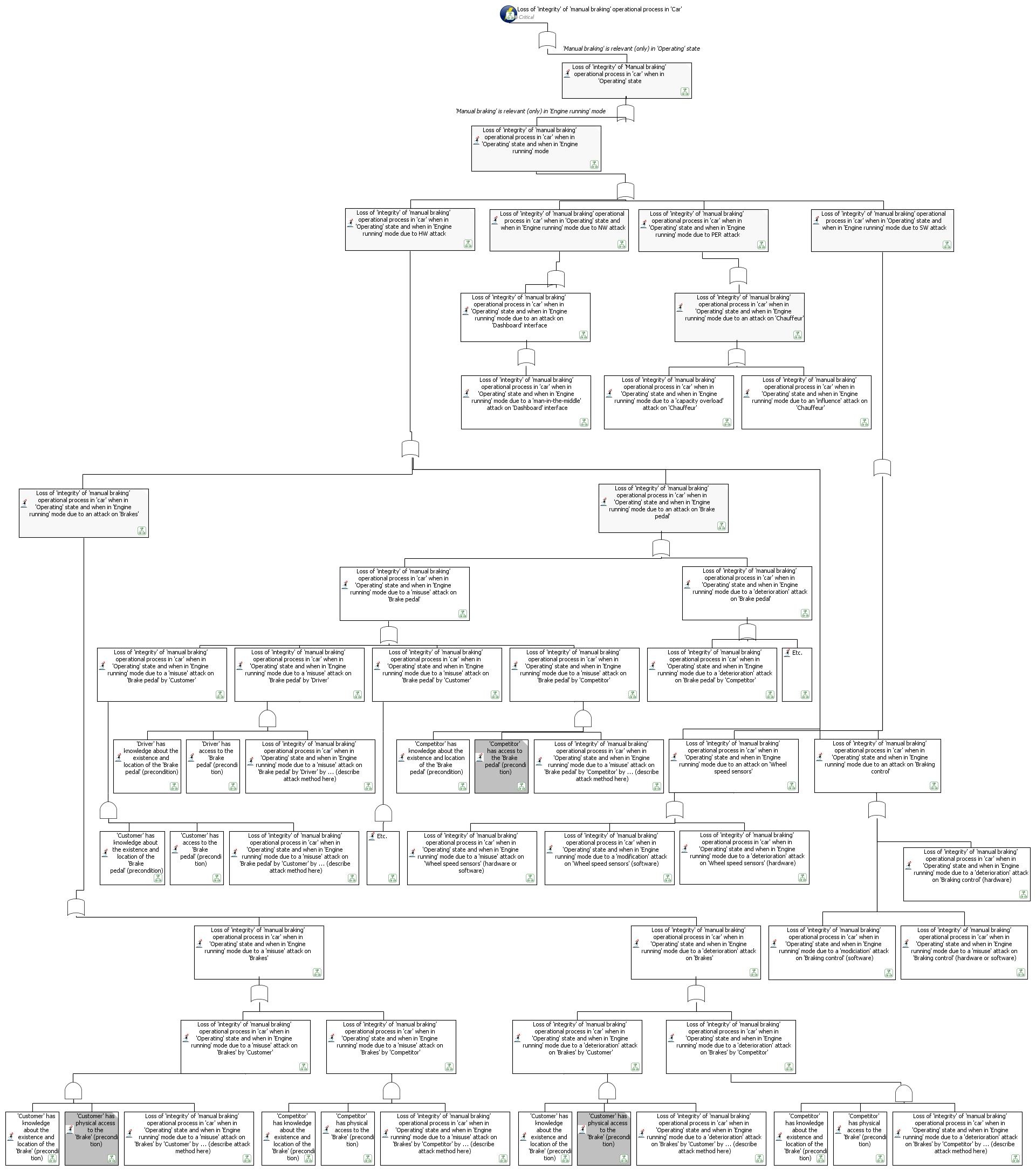}
\caption{Automatically (partially) constructed tree based on retained threat sources}
\label{FullTree}
\end{center}
\end{figure}

Figure \ref{FullTree} shows the result of the application of the algorithm on approximately half the attack nodes. The attack tree is not meant to be legible, but is shown here to give an idea of the complexity of the generated trees. In the Galileo risk assessment programme, some attack trees are known to stretch over 40 A4 pages. In our simple running example, the attack tree size is expected to more than double when the decomposition process is completed.

\section{Related Work}

There are few commercial products related to attack trees. The two most significant ones are SecurITree \cite{SecurITree} and AttackTree+ \cite{AttackTree}. None of them proposes support for the automated construction of attack trees.
 
The scientific community is very active on the construction of attack graphs \cite{ESC-TR-2005-054, Birkholz10}. However, attack graphs essentially focuses on attacker attempts to penetrate well-defined computer networks, rather than addressing socio-technical systems, especially when poorly or partially defined; and naturally, the complexity issues addressed by this community essentially relates to scaling to very large networks, rather than to usability, i.e. trying to construct an attack tree skeleton that can be easily reworked and maintained by human security experts.
 
Some research papers do address socio-technical systems. For example, \cite{Lamsweerde03fromsystem} and \cite{eemcs23000} propose formal approaches to generate attack trees, respectively based on system goals, and security policies. These studies can be seen as upstream complements, but they also require specific frameworks. Our approach attempts to base most of its attack tree extraction on an industrially used framework.

\section{Conclusion and Way Forward}

As can be seen from Figure \ref{FullTree}, significant \emph{draft} trees can be automatically generated, even with a simple case-study. Beyond saving security expert time to build the attack tree, a systematic approach is enforced, which ensures the completeness of the threat and vulnerability analysis. Moreover, the top-level structure of the trees is normalised throughout the study, thus easing the readability and understanding of the trees by third-parties; only the lower parts of the tree are left for manual completion. Last but not least, traceability to architecture artefacts is set-up at no additional cost, thus offering support for impact analysis when the system architecture evolves.

Our approach is foreseen for use only for large new systems, and not to support change in existing systems. The main reason for that is that one starts the design of a  large new system with a white sheet; in this setting, the approach allows for the rapid construction of some core attack trees, and it eases the impact management as the system's design evolves. When system delivery is concluded, the commercial contract(s) for system provisioning and system risk assessment usually terminate(s). A new commercial contract may now be enacted to support Security Information and Event Management (SIEM) at operation-time, possibly with a different industry. There are multiple approaches to SIEM, but most are based on attack graphs and / or Complex Event Processing (CEP) techniques, rather than on attack trees. Reuse of the knowledge captured in attack trees to feed the SIEM would definitively be useful if the same industry manages to win both the design-time risk assessment study and the SIEM contract, but we have not (yet) investigated this research path.

\begin{figure}[htbp]
\begin{center}
\includegraphics[width=7.5cm]{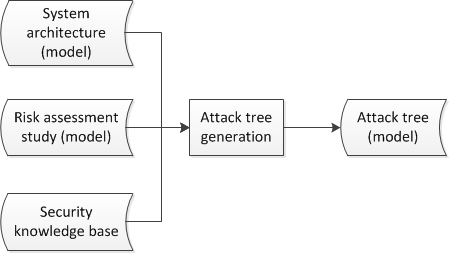}
\caption{Inputs and outputs of the attack tree generation approach}
\label{IO}
\end{center}
\end{figure}

To support the automatic attack tree generation, inputs are required from: (i) an architecture framework, in particular data from the operational architecture and the logical architecture; (ii) a risk assessment tool, in particular in terms of feared events, primary assets, supporting asset types, threat sources, etc.; (iii) a security knowledge base, in particular in terms of supporting asset types and threat caracterisation (cf. Figure  \ref{IO}).

Beyond the positive points stated above, this feasibility study has also highlighted some issues; further research has been shown to be required on a number of topics, in particular:
\begin{itemize}
\item the current work was focused essentially on operational processes and / or logical functional chains; further research is required to cope with attacks on data and communications between logical components;
\item the proposed algorithm has made the assumption that ports on logical component were tagged to specify if they provided logical versus physical access; this approach needs to be further validated because it requires additional work compared to current industrial engineering practices;
\item the proposed approach extensively analyses the logical architecture in search of the relevant attack entry points, rather than using the supporting assets, as defined the EBIOS \cite{ebios} risk assessment study; the consistency between attack entry points and supporting assets must be ensured;
\item the high-level structuring of the attack tree is based on states and modes; it has been implicitly assumed in this feasibility study that there were only two levels (i.e. a state-level and a mode-level which refines the states) in the system's architecture; however, it can be supposed that very complex systems may use sub-states and / or sub-modes; further research is required to assess the importance of considering more than two level state machines;
\item the approach fails to capture that some of the attacks may be led during the systems development life-cycle itself, e.g. theft of detailed design documents during the transmission of those documents between the development teams; to cover this type of attack, we may need to define a development life-cycle (e.g. specification, coding, integrationÉ) by opposition to a operation-time life-cycle, or, at minima, include in the attack tree an additional branch for attacks during the development phase, to be manually completed by the security expert; for certain programmes, it could also be interesting to include in the tree, attacks that could occur during system disposal;
\item the feasibility study has been run using the Thales Melody architecture framework; synthesising and generalising the required inputs from the architecture framework is required, in particular to assess connection possibilities to other, possibly commercial, architecture frameworks; formalising the algorithm would then be possible, for example using the Object Constraint Language (OCL) if the architecture is expressed in the Unified Modeling Language (UML);
\item logical {\tt AND} gates have been introduced in the tree only to cope with attack pre- and post-conditions; further work is required to deal with logical redundancy in operational and / or functional chains;
\item to ease the readability and understanding of the automatically generated tree, significantly long node labels have been used to precisely define the attack and its enactment conditions; since many attack tree tools, whether academic or commercial, do not support long node labels or restrict node labels to a unique line, further research is required to assess the trade-off between readability and scalability; tree layout may also be a constraint to consider, as opening a poorly-formatted large automatically-generated attack tree may be perturbing for the end-user;
\item attack tree node generation can sometimes be disputable; in those cases, we have always favoured node generation, so as to ensure the completeness of the analysis, making up for possibly unrealistic node generation by using node annotations or colour codes to attract the security experts' attention; this approach, originating from expert judgement, needs to be further validated;
\item the threat source selection algorithm is the most complex part of the overall approach, but still remains the least convincing; further research is required to simplify the approach.
\end{itemize}

\section{Acknowledgment}

The research leading to these results has received funding from the ITEA2 MERgE project. The help and support from engineers at Thales Communications \& Security have been of particular value.

\bibliographystyle{eptcs}
\bibliography{generic}
\end{document}